\documentclass[a4paper]{spie}  
\usepackage[]{graphicx}
\usepackage[]{amsmath}
\title{Grazing Incidence Reflection and Scattering of MeV Protons} 
\author{Bernd Aschenbach\supit{}
\skiplinehalf
\supit{}Max-Planck-Institut f\"ur extraterrestrische Physik, Giessenbachstrasse, 85741 Garching, Germany
}
\authorinfo{Further author information: (Send correspondence to B.A.)\\B.A.:
E-mail: bra@mpe.mpg.de, Telephone: +49 89 30000 3561}
\begin{document} 
  \maketitle 
\begin{abstract}
Treating protons as de Broglie waves shows that up to a few MeV energies protons experience 
total external reflection using the index of refraction concept for the target 
earlier applied to electrons. 
Angular scattering distributions can be explained by random surface scattering as known for X-rays. 
Applied to the {\it{Chandra}} and {\it{XMM-Newton}} X-ray telescopes the calculated 
reflection efficiencies can explain the observed degradation of the X-ray CCDs for both missions. 
Some discussion about the possibility of realizing imaging sub-MeV and MeV proton optics is presented.   
\end{abstract}
\section{INTRODUCTION}
\label{sec:intro}  
The X-ray astronomy observatory {\it{Chandra}} was launched into orbit in July 1999. Very early in the mission a performance degradation, predominantly 
with the front illuminated CCD detector ACIS-FI was noticed. Apparently some damage was done to the CCD which resulted 
in degradation of spectral performance due to an increase of charge transfer inefficiency (CTI)$\sp 1$. 
Investigation of the problem revealed that the most likely source for the damaging were protons with energies between 
100 keV and a few MeV$\sp 2$.    
Analysis also showed that the only path for protons getting to the focal plane is through the mirror system of the 
X-ray telescope as the structure has more than sufficient shielding against such protons. 
 The X-ray astronomy observatory 
{\it{XMM-Newton}} at that time was prepairing for the flight readiness review (FRR) in October 1999, and the team was 
badly concerned because the observatory carrys similar CCDs.
Meanwhile measures have been taken to prevent the protons reaching the CCDs, like moving the device out of the 
focal plane ({\it{Chandra}}) or cover it with a solid absorber implemented on the filter wheel in front 
({\it{XMM-Newton}} EPIC), 
where possible. 

The scattering of protons off the mirror reflecting surfaces came totally unexpected, and the question is 
what physical process is at work. Rutherford scattering was considered because of the very pronounced 
forward scattering, and scattering by an angle of something like 0.5$\sp{\circ}$ to 1$\sp{\circ}$ is just needed.   
However, for a proton of 100 keV energy to get scattered by 0.5$\sp{\circ}$ it needs an electrical field of 
about 900 V. This is equivalent to the binding energies of the electrons in the N-shells of a Au atom, whereby 
Au is the coating of the {\it{XMM-Newton}} mirrors. Most of the interatomic space is found outside of the 
N-shell region. Therefore it is much more likely that the scattering angle is realized by multiple scattering 
processes in the coating. 
Computer based Monte-Carlo routines like TRIM and GEANT4 have been available but had to be  
substantially modified to simulate multi-scattering
of a charged particle in matter under grazing incidence$\sp{2, 3}$.   
The dedicated measurements of proton scattering off an {\it{XMM-Newton}} representative flat mirror sample 
undertaken by Rasmussen et al.$\sp{4}$ are quite well reproduced by the GEANT4 simulations$\sp{4}$ within a 
factor of 2 to 3 as far as the efficiency is concerned. I note here that the telescopes employ two reflections 
or scatterings, which might increase the difference to some larger level. 
Details of the scattering distributions are not reproduced like a turn down of the efficiency towards low incidence angles. 
Measurements show a maximum at some scattering angle for fixed incidence angle$\sp{4}$. However, at the very 
low angles the measurements might have had some problems, so that more detailed studies are needed to confirm the trend.          

In this paper I would like to present a totally different approach to the proton scattering phenomenon using the dualism 
between particle and wave, treating the proton as a wave. Much of the material has been presented at several {\it{XMM-Newton}} emergency meetings in October 
1999 prior to FRR. 

\section{Proton reflection}

In 1924 Louis de Broglie postulated that any moving particle or object had an associated wave with wavelength 
${\lambda}$ = h/p.
h is Planck's constant, p is the momentum of the particle. 
In 1927 Davisson and Germer turned the postulate into reality, when they observed diffraction patterns of electrons 
reflected off the surface of crystallized nickel. The diffraction patterns also showed that the target had to have an 
index of refraction, which is slightly greater than unity for electrons. 
Subsequently the index of refraction was explained by introducing the concept of an inner potential. 
This is illustrated by Eqs. (1) to (6).

\begin{equation}
        \label{eq:phasev_a}
\rm{{c\sb a}} = {\frac{\rm E}{\rm{{p\sb a}}}}
=
{\frac{\rm E}{\sqrt{\rm{{2~m~E}}}}}
        \end{equation}

\begin{equation}
        \label{eq:mom_i}
\rm{{p\sb i}} = {\sqrt{\rm{{2~m~(E~-~U)}}}}
        \end{equation}

\begin{equation}
        \label{eq:phasev_i}
\rm{{c\sb i}} = {\frac{\rm E}{\sqrt{\rm{{2~m~(E~-U)}}}}}
        \end{equation}

\begin{equation}
        \label{eq:ind_refr}
\rm{{\mu}} = {\frac{\rm{c\sb a}}{\rm{{c\sb i}}}}
=
\sqrt{\frac{(E~-~U)}{\rm{ E}}}
        \end{equation}

\begin{equation}
        \label{eq:int_ener}
\rm{U} = \rm{{Z~e~V\sb 0}}
\end{equation}

\begin{equation}
        \label{eq:ind_ref_2}
\rm{{\mu}}
=
\sqrt{\rm{{1~-~Z~{\frac{V\sb 0}{V}}}}}
        \end{equation}

The particle's phase velocity in vacuum ${{\rm{c\sb a}}}$ is expressed by its energy 
${{\rm{E}}}$ and its momentum ${{\rm{p\sb a}}}$ and mass ${{\rm{m}}}$ (Equation 1). 
Penetrating the target the particle's momentum is changed to ${{\rm{p\sb i}}}$  because 
of some inner potential energy ${{\rm{U}}}$ (Equation 2), and so does the phase velocity ${{\rm{c\sb i}}}$  
(Equation 3). The index of refraction $\rm{{\mu}}$ is defined as the ratio of the  
two phase velocities (Equation 4). Replacing the energies by the corresponding potentials the 
index of refraction is given by Equation (6).  ${{\rm{V\sb 0}}}$ is called the inner potential,  ${{\rm{Z}}}$ 
is the particle's charge. For electrons ${{\rm{Z}}}$ = -1, and since $\rm{{\mu}}~>~1$, 
${{\rm{V\sb 0}}}~\ge~0$. In this way ${{\rm{V\sb 0}}}$ has been measured for quite a number of materials, 
and ${{\rm{V\sb 0}}}$ varies between 13 V and 17 V for Ni, Pb and Ag$\sp 5$. There appears to be a tendency that 
${{\rm{V\sb 0}}}$ decreases slightly with energy.  

For particles with ${{\rm{Z}}}~>~0$, for instance a proton, but it may be a positron or a muon or a 
positively charged ion,  $\rm{{\mu}}~<~1$. 
As for X-rays, for which also $\rm{{\mu}}~<~1$, total external reflection up to a maximal grazing angle 
${{\alpha}\sb t}$ must occur. Using Snell's law Eqs. (7) and (8) are derived. 

\begin{equation}
        \label{eq:snellius}
\rm{cos({\alpha}\sb t)} = {{\mu}}
         \end{equation}

\begin{equation}
        \label{eq:tot_ref_ang}
\rm{{\alpha}\sb t} =
\sqrt{\rm{{2~(1~-{\mu})}}}
        \end{equation}

For a proton of 100 keV energy or ${{\lambda}}$ = 9.1$\times$10$\sp{-12}$ cm, ${{\alpha}\sb t}$ 
= 42 arcmin using V$\sb 0$ = 15 V. This total reflection angle is about the same as the one for X-rays of a few keV energy. 

The complex index of refraction can be written as

\begin{equation}
        \label{eq:index}
\rm{n~=~1~-~\delta~+~i\cdot\beta}
         \end{equation}

An index of refraction is complete only with its imaginary part $\beta$,
 which for X-rays describes the absorption in matter. 
I adopt the same relation for protons, such that 

\begin{equation}
        \label{eq:imagin}
\rm{\beta~=~{\lambda}/(4~\pi~z)}
         \end{equation}

with \rm{z} being the proton penetration depth.
Using Eqs. (6) and (9) we get 

\begin{equation}
        \label{eq:real}
\rm{\delta~=~{\frac{1}{2}}~{\frac{Z~e~V\sb{0}}{E}}}.
         \end{equation}

For positively charged particles with Z$\ge$1, $\delta\ge$0, and vice versa. 
  
Like for X-rays the reflectance of protons is calculated using the Fresnel equations for an infinitely thin boundary. 
Fig. (1) shows the change of reflectance with incidence angle for some proton energies. As expected 
the reflectance can be very high at low incidence angles and then smoothly drops to high angles. At energies as low as 
10 keV more than 20$\%$ reflectivity is achieved at an incidence angle as high as 4$\sp{\circ}$.

   \begin{figure}
   \begin{center}
   \includegraphics[bb=46 68 439 660,height=12cm,angle=-90,clip=]{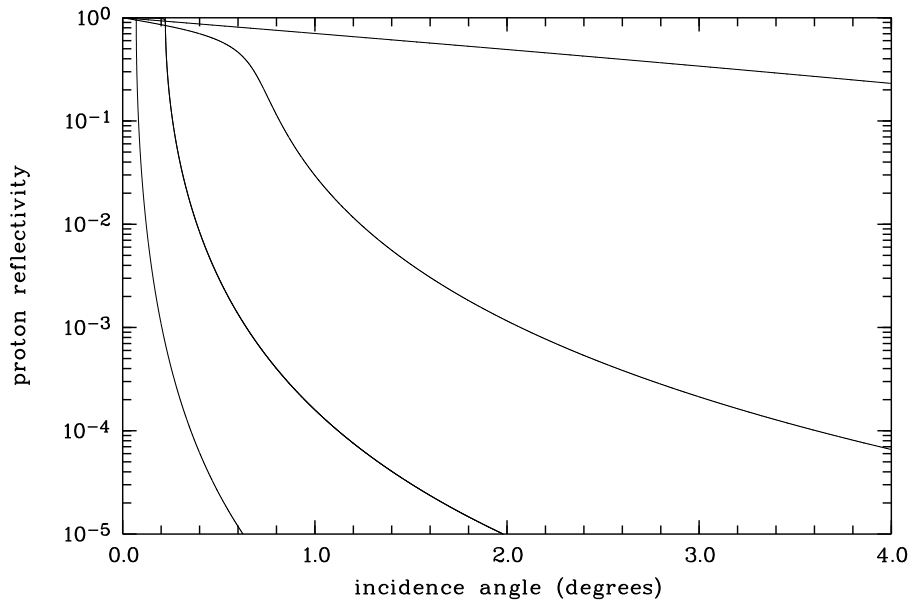}%
   \end{center}
   \caption[refl]
   { \label{fig:refl}
Reflectance of protons with energies of 10 keV, 100 keV, 1 MeV and 10 MeV, from the 
top right to the lower left, respectively, 
versus grazing incidence
angle for a gold coated mirror.}
   \end{figure}

Because of the similarity of the reflective behaviour of X-rays and protons at their proper energies, 
 and the index of refraction given, I have used ray tracing 
of both the {\it{Chandra}} and the {\it{XMM-Newton}} telescopes to compute their effective collecting areas 
for protons as a function of energy. Since the source of protons is diffuse and without angular structure, the 
source extent has been used to be {{2}}${\pi}$. The grasp of the telescope 
is defined as proton surface number density in the focal plane per incident  fluence unit, such that 
the product of the actual proton fluence impinging on the full telescope aperture in orbit times the grasp figure 
provides the proton count 
rate per detector area unit. 
For the RGS on {\it{XMM-Newton}} a third reflection is included which takes into account the reflection off the 
reflection gratings at an angle of 1.8${\sp{\circ}}$. Results are shown in Fig. (2).

   \begin{figure}[h]
   \begin{center}
   \includegraphics[bb=46 68 439 660,height=12cm,angle=-90,clip=]{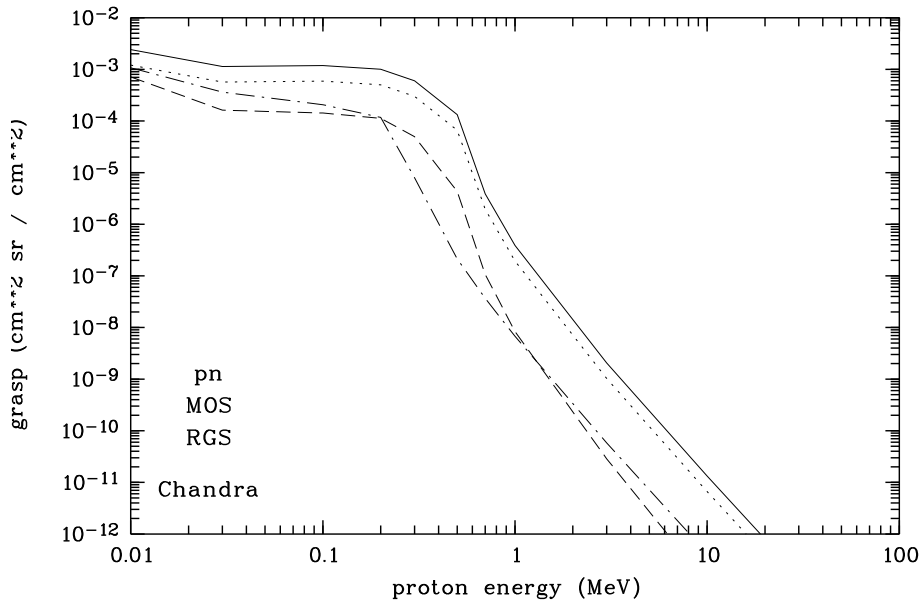}%
   \end{center}
   \caption[grasp]
   { \label{fig:grasp}
Energy dependence of the proton grasp for the Chandra telescope (dot-dash)  and the XMM-Newton telescopes
(solid line for the pn; dotted line for one MOS; dashed line for one RGS, the reflection off the grating has
been included).}
   \end{figure}

The highest grasp is obtained with the {\it{XMM-Newton}} telescope furnished with the pn-CCD in focus. Slightly lower, by about
a factor of two, comes the {\it{XMM-Newton}} telescope with a MOS-CCD in focus. The reduction is due to the obstruction
by the RGS gratings. The third reflection off the RGS gratings brings the grasp down by another factor of three at low energies and
more than a factor of 15 at energies above 0.5 MeV. The grasp of the {\it{Chandra}} telescope is slightly higher than the
RGS combination at energies below 100 keV, significantly higher in the 0.25  to 0.7 MeV range. This is attributed to the
different grazing angles and the number of reflections involved.

\section{CCD damaging and their lifetimes}

Damage to the X-ray CCDs occurs when protons dump energy in the charge transfer region, creating traps, which increase the CTI, 
degrade the energy resolution and ultimately render the CCD functionless when 
a "lethal" dose has been reached. This happens at a load of about 10$\sp{9}$ keV cm$\sp{-2}$ (L. Str\"uder, private 
communication). 
Above the charge transfer region , i.e. towards the incoming X-rays or protons, 
the CCDs have covering layers of Si and SiO$\sb{\rm 2}$, the thickness of which is different for the various types of CCDs used. 
The ACIS-FI, or more generally the MOS based CCDs, show the thinnest layers, whereas the pn-CCD has a very thick layer, 
for numerical values of the thicknesses see Table (1). 
Protons have to have a minimum energy to traverse the top layers and to reach the charge transfer region.

Fig. (3) shows the penetration depth of protons as function of energy. Column 3 of Table (1) displays these critical 
energies (E$\sb{\rm{crit}}$). They run from 160 keV (ACIS-FI) to 6 MeV for the pn. 
It is interesting to note that the X-ray telescope mirrors exhibit grazing angles at which these MeV or sub-MeV 
protons experience total external reflection. Though designed for imaging X-rays sources these mirrors are proton 
telescopes as well. How good their imaging rather than collecting properties are remains to be seen.

   \begin{figure}[h]
   \begin{center}
   \includegraphics[bb=46 68 439 660,height=12cm,angle=-90,clip=]{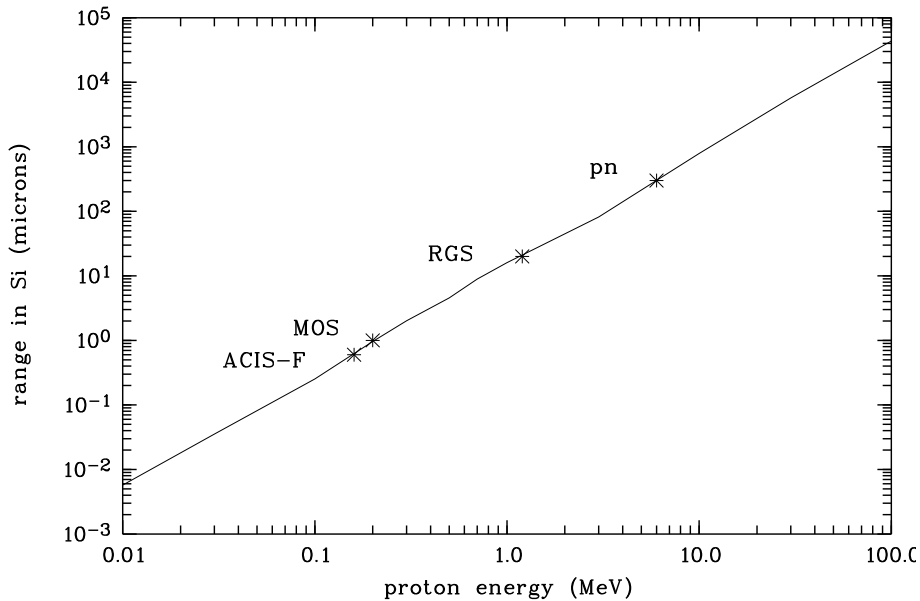}%
   \end{center}
   \caption[penetration]
   { \label{fig:penetration}
Energy dependence of proton penetration depth in Si. Asterisks denote the Si thickness and the corresponding minimum energy
for protons to reach the charge transfer regions of various X-ray CCDs.}
   \end{figure}

\begin{table}[h]
\caption{In October 1999 predicted orbital lifetimes of unprotected X-ray CCDs against proton damage.}
\label{tab:life}
\begin{center}
\begin{tabular}{|l|c|c|c|c|c|l|} 
\hline
\rule[-1ex]{0pt}{3.5ex}  instrument &Si depth& E$\sb{\rm{crit}}$&grasp(E$\sb{\rm{crit}})$&N$\sb p$(E$\sb{\rm{crit}}$)&
                                    on-CCD protons&lifetime  \\
                                    &(${\mu}$\rm m)& (MeV)&(cm$\sp 2$ sr/cm$\sp 2$)&(cm$\sp{-2}~\rm{sr}\sp{-1}~
                                    \rm s\sp{-1}$)&(cm$\sp{-2}~ \rm s\sp{-1})$& \\
\hline
\rule[-1ex]{0pt}{3.5ex}  ACIS-FI&0.6 &0.16&1.3$\times$10$\sp{-4}$&2.1$\times$10$\sp{5}$&27&3 days   \\
\hline
\rule[-1ex]{0pt}{3.5ex}  MOS&1.0 &0.2&5.$\times$10$\sp{-4}$&1.3$\times$10$\sp{5}$&65&1 day   \\
\hline
\rule[-1ex]{0pt}{3.5ex}  RGS&20. &1.2&3.5$\times$10$\sp{-9}$&2.5$\times$10$\sp{3}$&9$\times$10$\sp{-6}$&$\gg$10 years \\
\hline
\rule[-1ex]{0pt}{3.5ex}  pn&300.&6.0&1.2$\times$10$\sp{-10}$&6.5&8$\times$10$\sp{-10}$&$\gg$10 years   \\
\hline
\end{tabular}
\end{center}
\end{table}

Multiplying the grasp with the incident proton fluence gives the on-CCD proton brightness, listed in
column 6 of Table (1). The fluence has been taken from an Estec document$\sp{6}$ and represents the conditions in the
{\it{XMM-Newton}} type of orbit. Because of lack of better 
knowledge the same proton fluence has been applied to ACIS-FI. Dividing
the "lethal" dose by the on-CCD proton brightness a measure for the functional lifetime of the CCDs is estimated and
listed in the last column of Table (1). According to these estimates the functional lifetime of the ACIS-FI would have been
just a few days, the MOS CCDs would have degraded very quickly within a day or so. These numbers explain the
observed rapid degradation of the ACIS-FI; without the knowledge of the ACIS-FI problems and their explanation
the protection measures against sub-MeV protons for the {\it{XMM-Newton}} MOSs must have been taken much faster than for
{\it{Chandra}} when in orbit. Clearly the CCD used as detector in the RGS and the pn-CCD are pretty safe.

\section{Scattering and imaging - towards a proton telescope?}

So far, the proton reflection has been treated as being specular. Of course there will be scattering, both volume scattering, 
if the penetration depth is large, and surface scattering, if the penetration depth is short 
or the proton energy is low. 
For the surface scattering one might use the same formalism as for X-rays. 
This is described in Eqs. (12) and (13). The total integrated scatter $\rm{{TIS}}$ for angles larger than the 
scattering angle ${\varphi}$ increases with the surface roughness ${\sigma}$ for an incidence angle 
${\alpha}$ and wavelength ${\lambda}$. 
The imaging quality of the mirror depends on the width of the point spread function and the ${\rm{TIS}}$, which results  
from randomly distributed surface structures.
Because of the short proton wavelength, ${\rm{{TIS}}}$ is very large for reasonable values of ${\alpha}$.
For ${\rm{{TIS}}}$ = 0.5 (equivalent to a resolution half energy 
width (HEW)) and ${\alpha}$ = 0.5$\sp{\circ}$,  ${\sigma}<$4 m{\AA} for a proton of 200 keV. 
But microroughness is a band limited quantity and related to the spatial frequency distribution of the 
surface (c.f. Equation (13)). For example, if the spatial resolution of the mirror is set 
to 1 arcmin HEW (${\varphi}$ = \rm{{0.5 arcmin}}), the requirement of ${\sigma}<$ 4 m{\AA} is to be fulfilled 
for all spatial wavelengths \rm{{d}} shorter 
than 25 nm, i.e. over a microscopic scale. Over such a spatial scale it does not seem to be impossible to 
reach the required microroughness of a few m{\AA}.
Another aspect of imaging is
the contrast, which is defined by the height of the central peak of the image. If one accepts a low contrast image but with a central
prominent peak which
contains just 0.1$\%$ of the total flux, the requirement on ${\sigma}$ is reduced  to about 15 m{\AA}
over the same spatial frequency range. 

\begin{equation}
        \label{eq:tis}
\rm{{TIS}} = \rm{{\left[
\frac{4~{\pi}~{\sigma}~{\alpha}}
{{\lambda}}\right]\sp 2}}
        \end{equation}

\begin{equation}
        \label{eq:diff_ang}
\rm{{\varphi} = {\frac{{\lambda}}{d~{\alpha}}}}
        \end{equation}

\begin{equation}
        \label{eq:roughness}
\rm{{{\sigma}\sp 2 = {\sigma}\sb 0\sp 2~{\left[\frac{d}{d\sb 0}\right]\sp{{\gamma}~-~1}}}}
         \end{equation}

\begin{equation}
        \label{eq:diff_ang_1}
\rm{{{\varphi} = \left[\frac{\left(4~{\pi}~{\sigma}\right)\sp 2}{q}\right]
\sp{\frac{1}{{\gamma}~-~1}}~\frac{1}{d\sb 0}~\left(\frac{{\alpha}}{{\lambda}}\right)
\sp{\frac{3~-~{\gamma}}{{\gamma}~-~1}}}}
         \end{equation}

For lower spatial frequencies geometric slope errors dominate the point spread function$\sp{7}$. 
By X-rays the range of spatial frequencies 
of well polished mirrors has been explored down to scales of millimeters, shorter scales are fairly unexplored, except of upper 
limits of the microroughness up into the few millimeter spatial wavelength region.
The region of shorter spatial wavelength scales is largely unexplored, but leaving polished mirrors aside, microroughness 
values of the size required 
may be reached on single crystals, which may actually 
be of macroscopic size.
A double crystal arranged in a Wolter type telescope configuration may actually provide an imaging device.

Present day mirrors are likely to have much larger surface roughness in the sub-mm spatial 
wavelength scales (this is not an issue for high resolution X-ray mirrors), and 
their point spread functions are expected to  be fairly broad 
due to the strong scattering. 
One can use Eqs. (12) and (13) together with Eqs. (14) and (15) to make a rough estimate about the HEW
width of an up-to-date polished mirror. For such an estimate one has to make an assumption about 
the power spectral density ({\rm{{psd}}) as function of the spatial frequency {\rm{f}}. Assuming  
{\rm{{psd}}} $\propto$ \rm{{f}}$\sp{-{{\gamma}}}$ with \rm{{d}} $\propto$ \rm{{f}}$\sp{-1}$, 
${\sigma}\sp{2}$ reads as in Eq. (14). Equation (15) then provides the scatter angle beyond which the 
\rm{{TIS}} equals \rm{{q}}. 
As a numerical example one might take ${\gamma}~=~2$, \rm{{E = 200 keV}}, 
${\alpha}$ = \rm{{1$\sp{\circ}$}}, \rm{{q = 0.2}} and a fairly well polished mirror, available in the lab, with 
${\sigma}$ = \rm{{15}} \rm{\AA} and \rm{{d$\sb 0$ = 3 cm}, ${\varphi}$~\rm{$\approx~\pm{1}\sp{\circ}$}.
Such wide scattering distributions are likely to be expected from polished mirrors and their metallic coatings.
For the best polished mirrors with ${\sigma}$ = \rm{{3}} \rm{\AA} and \rm{{d$\sb 0$ = 3 cm}, 
${\varphi}$~\rm{$\approx~\pm{2.3}$ arcmin}. 
But it is unknown whether the psd continues down into the 
micron range 
with the power law chosen. Any residual microroughness will widen the point spread function with long tails in the 
scattering distribution.

Shortly before launch of {\it{XMM-Newton}} Rasmussen et al.$\sp 4$ have made proton scattering measurements with 0.3 MeV, 0.5 MeV and 
1.3 MeV protons using a representative mirror sample. The measurements were done by scanning over the incidence angle 
at three fixed scattering angles, from which proton efficiency and scattering distribution were deduced. 
For all three scattering angles the peak of the scatter distribution was within 5 arcmin of twice the incidence angle, 
which is expected from Snell's law 
and pronounced specuclar rflection. 
A coarse estimate of the scatter profile can be obtained from the 
1.3 MeV measurements at an incidence angle of $\alpha$ =0.7$\sp{\circ}$, which amounts to $\varphi$= -0.35$\sp{\circ}$ to 
$\varphi$= 1$\sp{\circ}$, i.e. a FWHM of 1.35$\sp{\circ}$. The minus sign means scattering towards the mirror, and the plus sign 
stands for scattering away from the mirror. At 0.3 MeV the width of the scattering distribution away from the mirror is 
similar. 
These experiments seem to confirm the proton wave model using Snell's law, and produce a wide scattering distribution, as 
expected from surface scattering. The asymmetry in the scattering distribution with respect to the direction towards and 
away from the mirror at very shallow incidence has been observed in X-rays and is easily explained in the 
surface scattering theory.

\section{Conclusion}

Describing the proton as de Broglie wave in conjunction with an  
index of refraction less than one, predicts total external reflection of protons. 
Using the values for the inner potential 
observed in electron measurements allows, together with the Fresnel equations, computing of  
reflectance values. Total reflection at incidence angles of less than one degree occurs for protons with 
energies less than a few MeV. Ordinary targets, even fairly well polished mirrors, are likely to have 
surface mircroroughness values which lead to scattering distributions one degree or more wide. Targets 
with much lower microroughness for spatial wavelengths in the sub-micron region are needed, which might be 
realized with crystals or other materials, to use this effect for building proton optics or a proton 
scatterometer, with which structures in the nanometer range might be revealed. The total external reflection occurs not 
only for protons but for any positively charged particle, including positrons, muons 
and ions. 
In principle, there is no limit to the particle energy for total external reflection inherent to the 
theory presented but some experimental practical limits are coming up with the total external reflection 
angle 
as small as one arcmin, which would correspond to an energy of about 175 MeV.

\section{Acknowledgments}
I would like to thank Andy Rasmussen for permission of using measurement results prior to publication, and 
Gerald Drolshagen for permission of making public use of the cited ESA report.
 

\section*{REFERENCES}

\begin{enumerate}
\item G.~Y. Prigozhin, S.~E. Kissel, M.~W. Bautz, C. Grant, B. LaMarr,
      R.~F. Foster, G.~R. Ricker,
      "Characterization of the radiation damage in the Chandra x-ray CCDs",
      {\it{Proc. SPIE}} {\bf{4140}}, pp. 123-134, 2000.
\item J.~J. Kolodziejczak, R.~F. Elsner, R.~A. Austin, S.~L. O'Dell, S. L., in 
      "Ion transmission to the focal plane of the Chandra X-Ray Observatory",
      {\it{Proc. SPIE}} {\bf{4140}}, pp. 135-143, 2000.
\item R. Nartallo, H. Evans, E. Daly, et al., 
      "Radiation Environment induced degradation of Chandra and implications for XMM",
      ESA report {\it{Esa/estec/tos-em/00-015/RN}}, 2000. 
\item A. Rasmussen, J. Chervinsky, J. Golovchenko, 
      "Proton scattering off XMM optics: XMM mirror and RGS grating samples",
      Columbia Astrophysics Laboratory, Document {\it{RGS-COL-CAL-99009}}, 1999
\item E.~W. Schpolski, 
      "Atomphysik I", VEB Deutscher Verlag der Wissenschaften, Berlin 1965
\item H. Evans, ESA report, {\it{esa/estec/wma/he/XMM/9}}, 1997 
\item B. Aschenbach, "Boundary between geometric and wave optical treatment of x-ray mirrors", 
      {\it{Proc. SPIE}} {\bf{5900}}, pp. 92-98, 2005.
\end{enumerate}
\end{document}